\documentclass{iau} 
\usepackage{graphicx}
\usepackage{caption}
\usepackage{subcaption}
\usepackage{vmargin}
\usepackage{fancyhdr}
\usepackage{amsmath}
\usepackage{amssymb}
\usepackage{array}
\usepackage{booktabs}
\usepackage{wrapfig}
\usepackage{float}

\title[Effect of Dense CSM on II-P SNe] %% give here short title %%
{Evolution \& Explosion of Massive Stars Leading to IIP-IIL SNe with MESA \& SNEC}

\author[Sanskriti Das \& Alak Ray]   %% give here short author list %%
{Sanskriti Das$^1$
 \and Alak Ray$^2$}

\affiliation{$^1$Department of Physics, Indian Institute of Technology Bombay, Mumbai- 400076, India \\ email: {\tt dassanskriti@gmail.com} \\[\affilskip]
$^2$Dept. of Astronomy \& Astrophysics, Tata Institute of Fundamental Research, Mumbai- 400005,India \\email: {\tt akr@tifr.res.in}}

\pubyear{2017}
\volume{331}  %% insert here IAU Symposium No.
\jname{SN 1987A, 30 years later}
\editors{A. Marcowith, M. Renaud, G. Dubner, A. Ray \& A. Bykov, eds.}
\begin{document}

\maketitle

\begin{abstract}
We show how the dense shells of circumstellar gas immediately outside the red supergiants (RSGs) can affect the early optical light curves of Type II-P SNe taking the example of SN 2013ej. The peak in V, R and I bands, decline rate after peak and plateau length are found to be strongly influenced by the dense CSM formed due to enhanced mass loss during the oxygen and silicon burning stage of the progenitor. We find that the required explosion energy for the progenitors with CSM is reduced by almost a factor of 2.    
\end{abstract}

\firstsection % if your document starts with a section, remove some space above using this command.

\section{Introduction}

\begin{figure}[]
\begin{subfigure}{0.5\textwidth}
\includegraphics[trim=2 0 13 5,clip, width =1.0\textwidth, height= 6 cm]{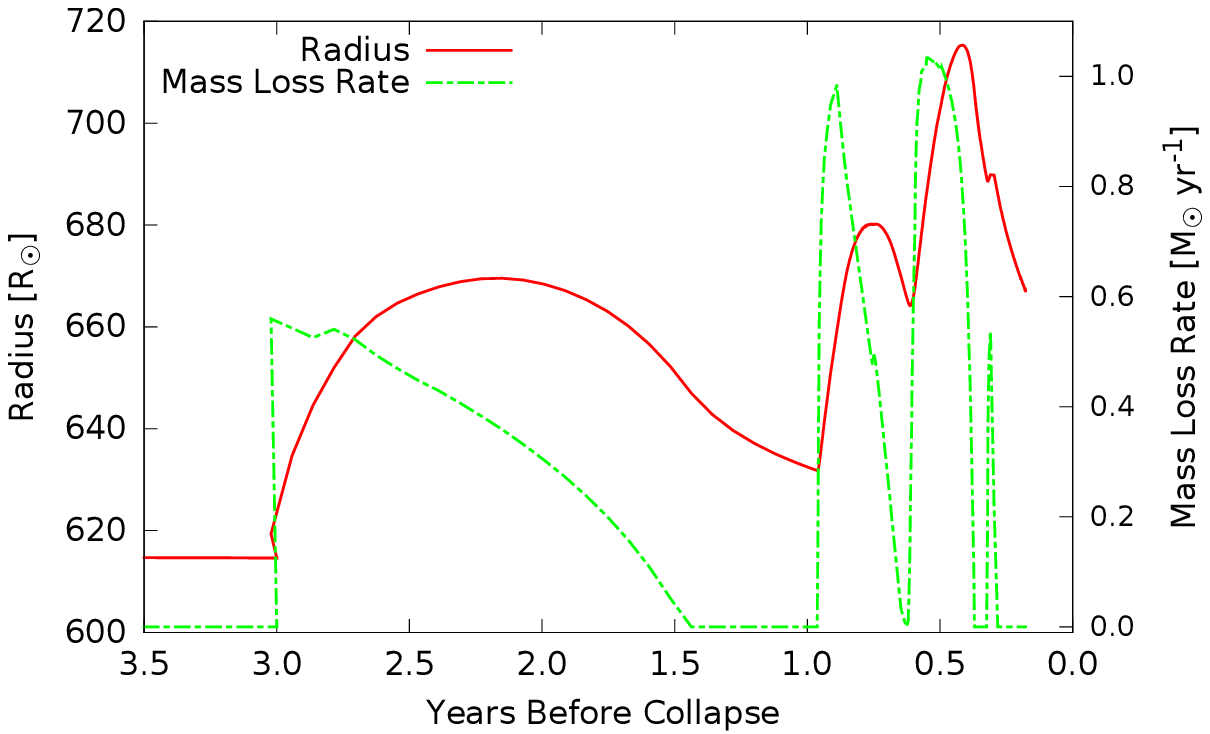}
\end{subfigure}
\begin{subfigure}{0.5\textwidth}
\includegraphics[trim=2 0 13 5,clip, width =0.95\textwidth, height= 6 cm]{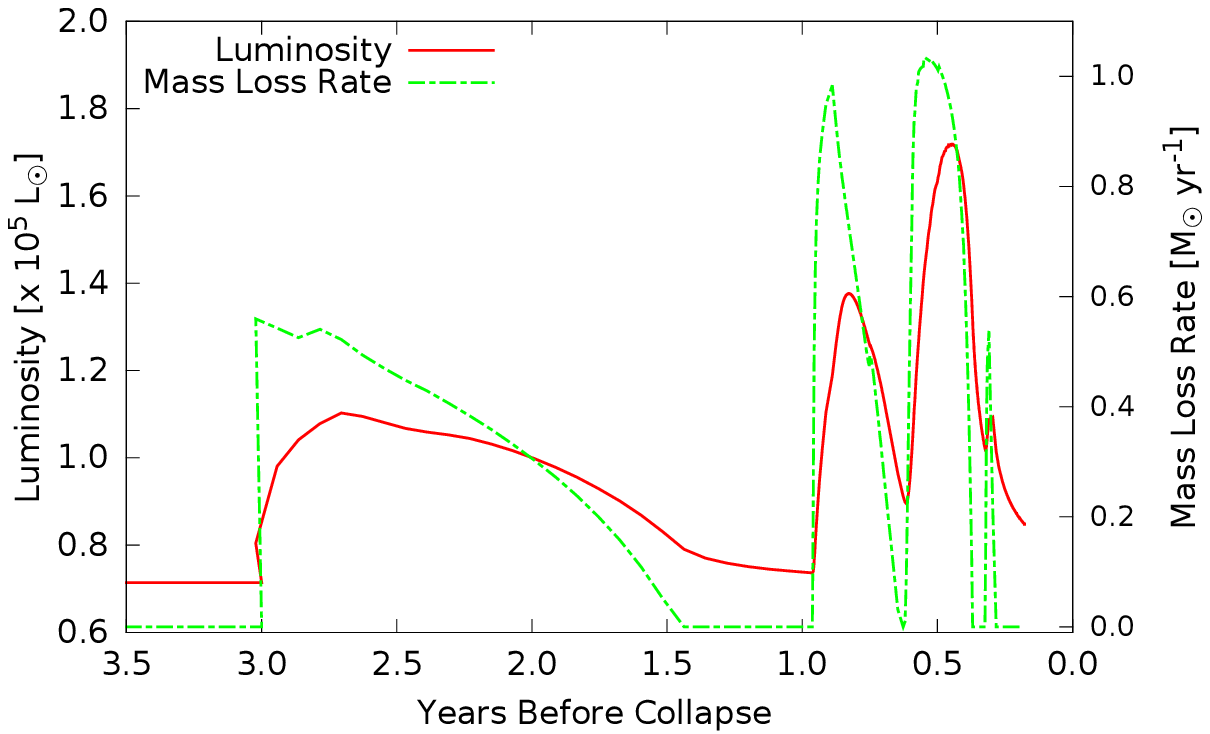}
\end{subfigure}
\caption{Variation in luminosity (right) and radius (left) with large mass loss rate in late stage of stellar evolution of a model of ZAMS mass 13 $M_\odot$.} 
\label{fig1}
\end{figure}

\hspace*{1 cm} Massive stars (ZAMS mass $> 8 M_\odot$) end up their lives through core-collapse supernovae (CCSNe). Type-II SNe, a class of CCSNe are identified by the P-cygni profile of hydrogen in their early spectra. Type II-P SNe cover a large fraction ($\sim 48\%$) of the whole type-II SNe population (\cite{SmithN_etal2011}). These have pronounced plateaus in their visible band light curves that remain within 1 mag of maximum brightness for an extended period e.g. 60-100 rest frame days and is followed by exponential tail at late times. II-Ls were originally separately classified by their brighter peak and linearly falling luminosity after peak, but \cite{Sanders_etal2015} shows that II-P and II-L form a continuous and statistically indistinguishable class of CCSNe. There has been search for a range of progenitors producing a continuous transition from II-P to II-L by arguing the qualitative similarity in their light curve pattern but so far it has not been possible to simulate II-L as an extreme case of II-P SNe (\cite{MorozovaV_etal2015}). Type II-n SNe show narrow emission lines of H-$\alpha$ due to the interaction of shock and the stellar ejecta with the circumstellar medium. So far, II-P/II-L and II-n have been kept into separate categories by virtue of their observed characteristics and inferred progenitor properties. Recently there has been considerations on the effect on the early light curves of II-P/II-L SNe due to dense circumstellar material immediately outside the progenitors (\cite{Valenti_etal2015}; \cite{MorozovaV_etal2017}; \cite{Yaron_etal2017}) and a subclass of \textit{moderately interacting SNe} has been identified. This may act as an intermediate of II-P/II-L and II-n and fill the gap between observed ZAMS (Zero Age Main Sequence) mass range of their respective progenitors(\cite{Moriya_etal2011}). \\ \\
\hspace*{1 cm} We have evolved the stars in MESA (Modules for Experiments in Stellar Astrophysics, \cite{Pax2011},\cite{Pax2013},\cite{Pax2015}) since their pre-ZAMS stage till the Fe core collapse with a history of enhanced mass loss rate in last few years, and constructed the CSM from that information. Our method is self-consistent in two ways: firstly, since the simulation of enhanced mass loss over a timescale of few years reveals the impact on surface properties (mainly luminosity and radius) naturally from the evolution (Fig. \ref{fig1}), the pre-SN progenitor carries the trace of the event with it and secondly, the modelling of the CSM also accounts for these changes in the star, and thus the CSM profile becomes more realistic. This is in contrast with the earlier works on the effect of CSM on II-P SNe where  artificially designed CSM profile is stitched to the pre-SN star independently modelled from stellar evolution codes (e.g. KEPLER) without any huge mass loss history (\cite{MorozovaV_etal2017}; \cite{Moriya_etal2017}). The progenitor with CSM is then exploded via SNEC (SuperNova Explosion Code)(\cite{MorozovaV_etal15}) and the light curve is compared against the light curve of the progenitor with no excess mass loss history and hence no dense CSM in the immediate vicinity of the exploding star. We make a comparative study of the models best fitting the optical light curve of Type II-P SN2013ej (sometimes described as II-L because of fast decline of 1.74 mag/100 days in intermediate stage) and show in next section how the models with CSM improves the fitting. \\ 
%\begin{wrapfigure}{l}{0.6\textwidth}
%\includegraphics[trim=2 0 10 5,clip, width =0.6\textwidth, height= 6.5 cm]{R_vs_Mdot.eps}
%\vspace{0.1em}
%\includegraphics[trim=2 0 10 5,clip, width =0.6\textwidth, height= 6.5 cm]{L_vs_Mdot.eps}
%\caption{Variation in luminosity and radius with large mass loss rate in late stage of stellar evolution of a model of ZAMS mass 13 $M_\odot$.} 
%\label{fig1}
%\end{wrapfigure}

\section{Circumstellar medium structure and explosion characteristics}
\begin{figure}[t]
\begin{subfigure}{0.5\textwidth}
\includegraphics[trim=0 0 14 8,clip, width =0.95\textwidth, height= 6 cm]{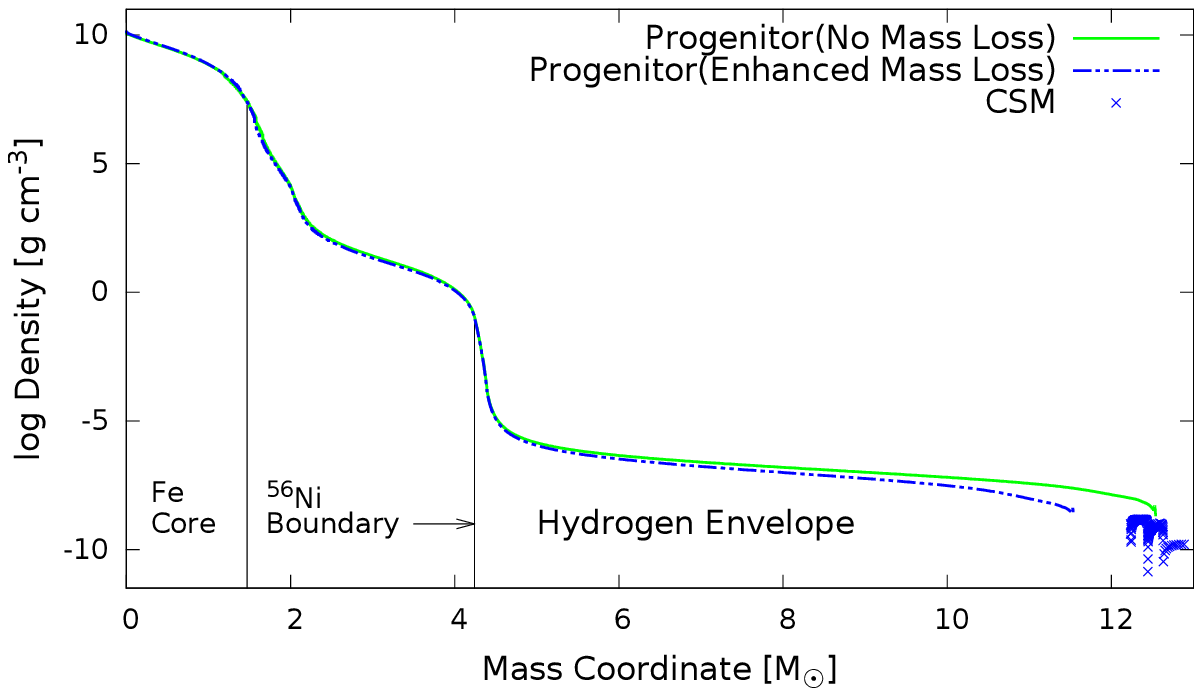}
\end{subfigure}
\begin{subfigure}{0.5\textwidth}
\includegraphics[trim=0 0 14 7,clip, width =0.95\textwidth, height= 6 cm]{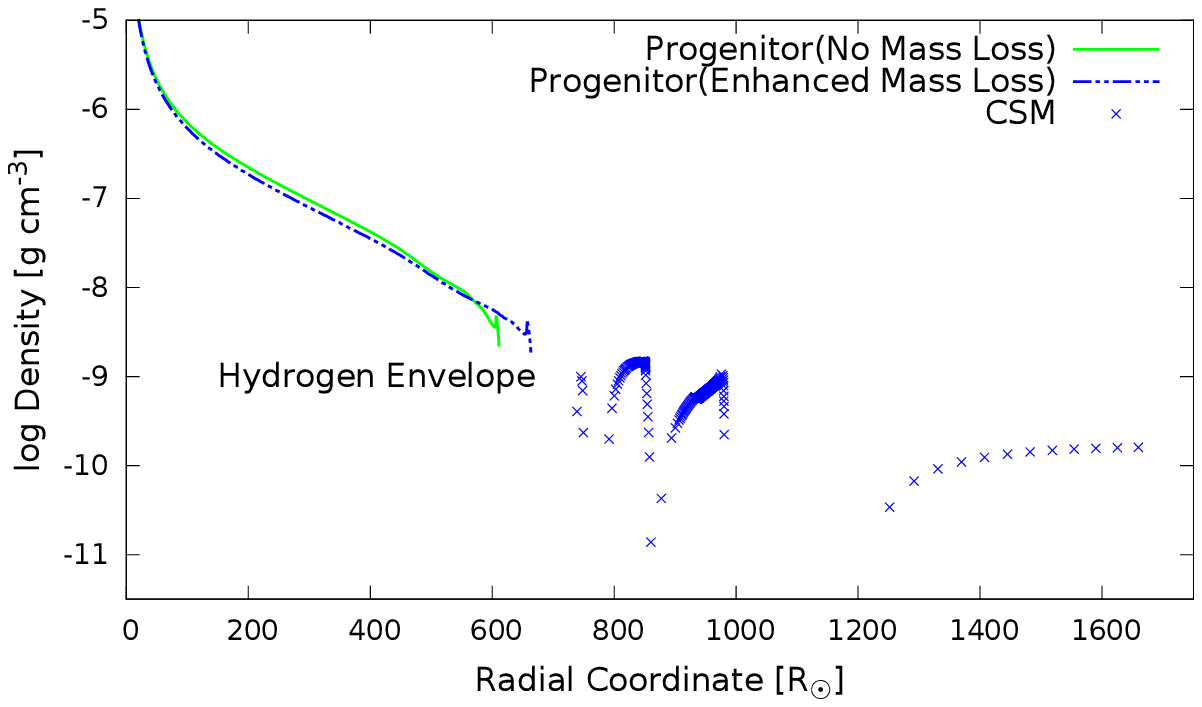}
\end{subfigure}
\caption{Pre-SN density profile of a progenitor of ZAMS mass 13 $M_\odot$. The model without enhanced mass loss and the one with shells of CSM have been shown in same plot for comparison. The radial profile of the central part of $\sim 4.5 M_\odot$ has been excluded from the plot on right side to show the CSM with better resolution.} 
\label{fig2}
\end{figure}

\begin{figure}[]
\begin{subfigure}{0.5\textwidth}
\includegraphics[trim=10 12 0 33,clip, width =0.985\textwidth, height= 6 cm]{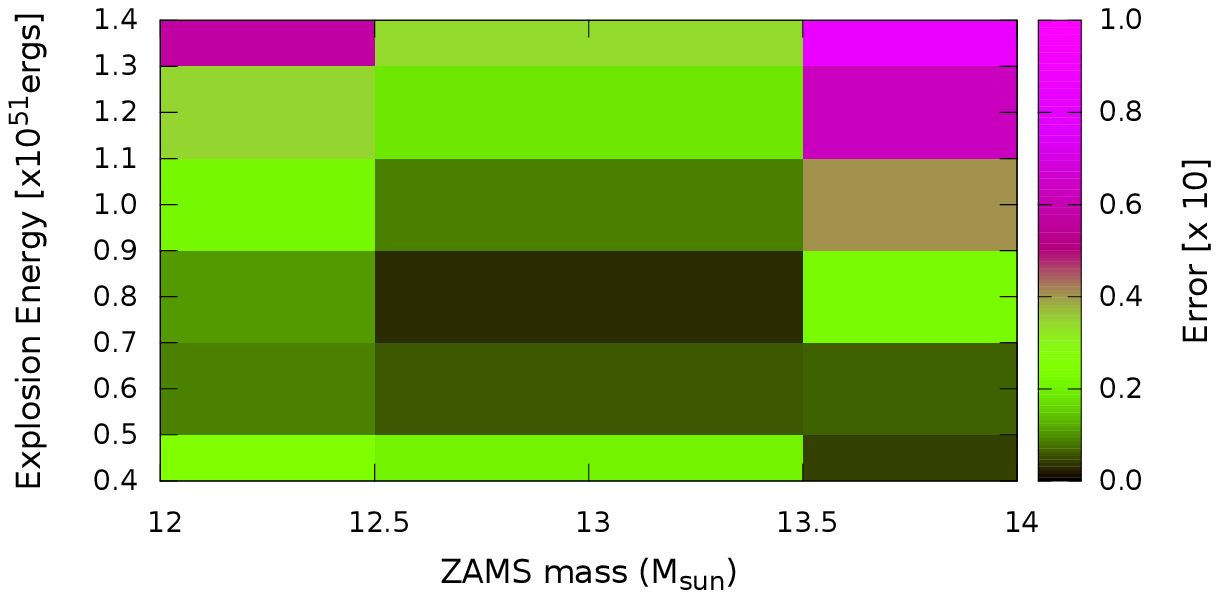}
\end{subfigure}
\begin{subfigure}{0.5\textwidth}
\includegraphics[trim=10 14 0 33,clip, width =0.985\textwidth, height= 6 cm]{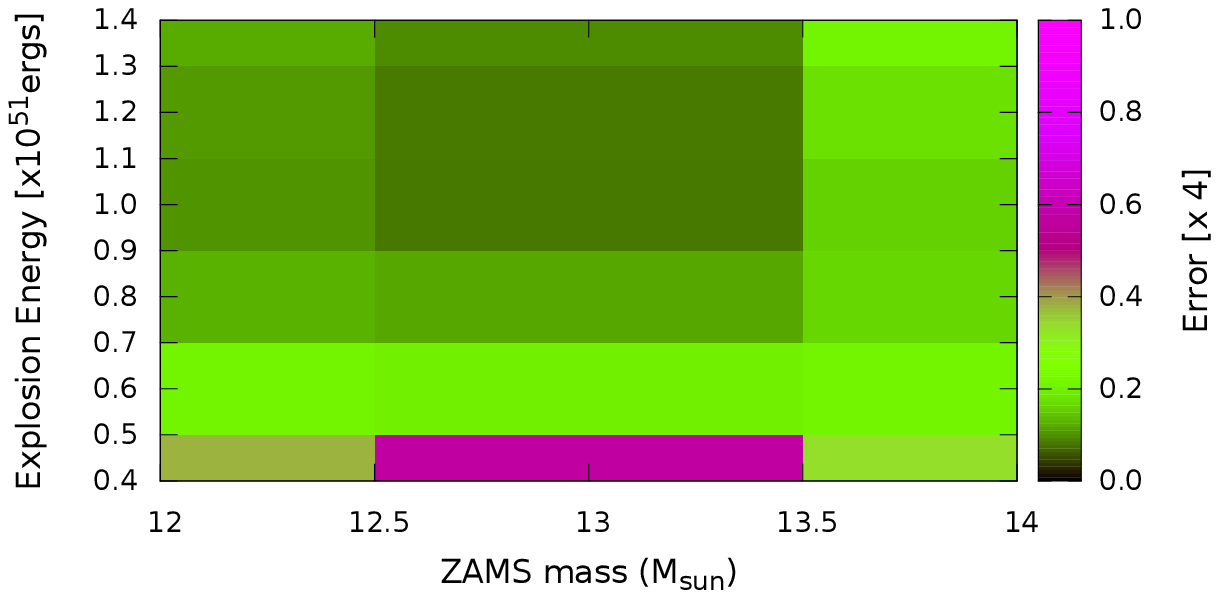}
\end{subfigure}
\caption{Error maps of the models with (left) and without (right) CSM of different ZAMS mass and explosion energy against V,R and I bands of SN2013ej (data: \cite{Yuan_etal2016}). The photometric error in observational data, errors in estimation of the distance of host galaxy and average galactic extinction (taken from \cite{Huang_etal2015}) have been included in the error analysis.} 
\label{fig3}
\end{figure}

\begin{figure}[h]
% \vspace*{-2.0 cm}
\begin{center}
 \includegraphics[trim=5 0 10 10,clip, width=15 cm, height= 10 cm]{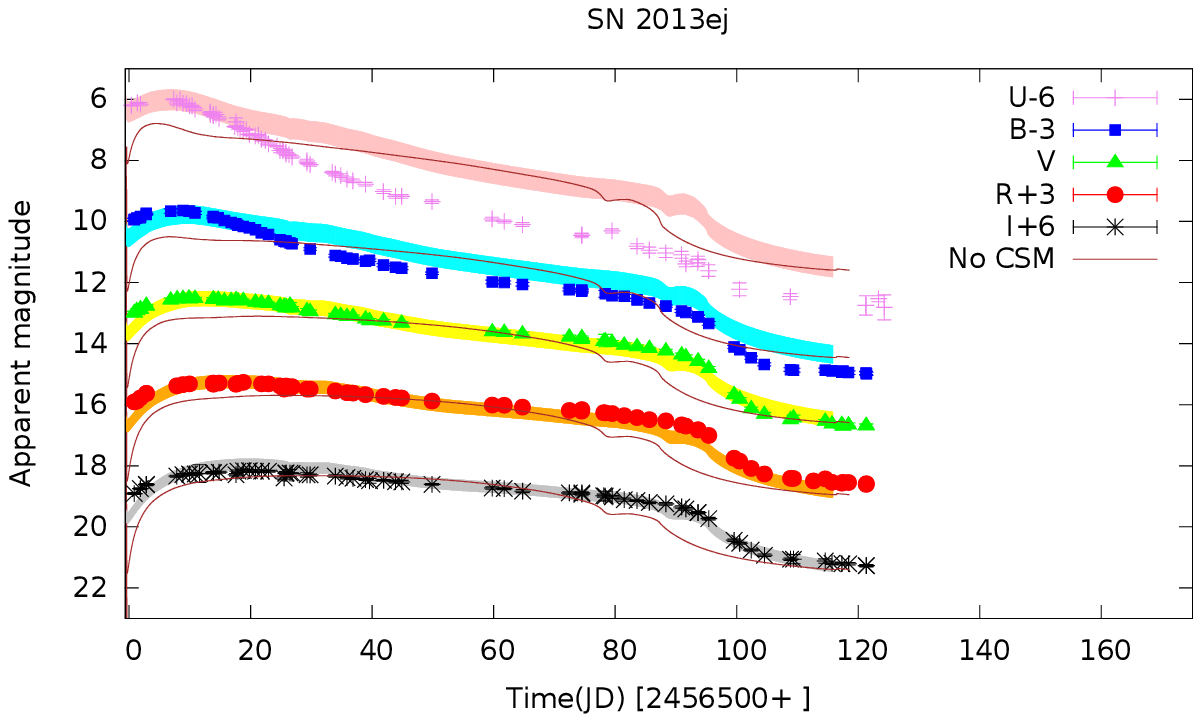} 
% \vspace*{-1.0 cm}
 \caption{The UBVRI light curves of SN2013ej (data: \cite{Yuan_etal2016}). The curves for best-fitted model (fitted against VRI bands till the plateau of $\sim$ 99 days) with CSM are shown with the errors at each point. Values of distance and extinction estimation have been taken from \cite{Huang_etal2015} [$d= 9.6\pm0.5$Mpc, $A_V=0.37\pm0.19$ with relative extinction relation borrowed from \cite{Cardelli}]. The best-fitted model without CSM is shown in thin lines for the sake of comparison.}
   \label{fig4}
\end{center}
\end{figure}

\hspace*{1 cm} To study the effect of CSM in the context of SN2013ej we have taken 0.295$Z_\odot$ metallicity, which is the equal to the metallicity of the nearby HII region number 197 of \cite{Cedres} and ZAMS mass range of 12-19 $M_\odot$ for all of our MESA models. We use Ledoux criterion for convection, and take mixing length parameter $\alpha_{mlt}$ = 2, exponentially decreasing overshooting (both for the core and the convective shells) with $f_{ov}$ = 0.025 and $f_0$ = 0.05, and semi-convection efficiency $\alpha_{sc}$ = 0.1. Near the center of the star we increase mass resolution with mesh$\_$delta$\_$coeff$\_$for$\_$highT = 1. During continuous mass ejection temporal resolution has been kept to default value with varcontrol$\_$target = $10^{-4}$. The resolution is increased by using max$\_$timestep = $0.05\times10^7$ seconds during the episodes of enhanced mass loss rate.The average mass loss rate has been kept to $\sim 10^{-6} M_\odot yr^{-1}$ (Vink's scheme for hot ($T_s > 10^4 K$) with wind $\eta$=1 and Dutch scheme for cool wind with $\eta$=1 and 0.5) constrained by X-ray studies which has probed mass loss rate in the range of 40-400 years before the explosion (\cite{Chak_etal2016}). %By comparing our models with pre-SN archival images taken 10 and 8 years before the explosion (\cite{Fraser_etal2014}) the ZAMS mass range is further constrained to 12-14 $M_\odot$.It is consistent with the upper limit inferred from nebular phase observation (\cite{Yuan_etal2016}) as well. 
The enhanced mass loss rate was triggered by hydrogen stripping (remove$\_$H$\_$wind$\_$mdot) and flashes (flash$\_$wind$\_$mdot) with an upper limit of $\sim 1 M_\odot yr^{-1}$ in last $\sim3$ years of evolution. Any heavy mass loss much before this would move away too far and become dilute to substantially affect the early light curves. Noting that the bolometric luminosity and radius vary with mass loss rate with certain correlation (Fig. \ref{fig1}), the CSM profile has been calculated from the history of mass loss rate, radius and surface temperature assuming asymptotic wind speed of 10 km/s. Although the launch velocity was found to be correlated with mass loss rate, we assume the ejected gas has had enough time to relax to the asymptotic speed during the episodic ejection. The composition in CSM is assumed to be the same as that of the surface. In contrast to a smooth continuous inverse square radial profile of density our constructed CSM is a collection of discrete shells of non-monotonically varying density. The resultant CSM has been stitched to the progenitor (e.g. Fig. \ref{fig2}) and exploded via SNEC using a thermal bomb. The explosion energy has been varied in the range of $ 0.4-1.4 \times 10^{51}$ erg. $^{56}$Ni mass has been kept fixed at 0.0207$M_\odot$ and evenly spread from outside the Fe core to the middle of He core (Fig. \ref{fig2}). After boxcar smoothing this distribution gets modified with a tail extending upto the He core boundary.The mass of the NS has been taken to be equal to Fe core mass. The light curves have been calculated till 120 days to cover a part of the radioactive tail after plateau. The magnitude in different bands have been calculated using proper bolometric corrections with blackbody approximation.\\ \\
\hspace*{1 cm} The resultant light curves in V, R and I bands have been compared with the data given by \cite{Rich2014} and \cite{Yuan_etal2016}.\footnote{Since SNEC does not calculate Fe line blanketing in U and B bands, these bands are excluded from goodness of fit analysis.} The reduced error of our SNEC outputs against the data has been defined following \cite{MorozovaV_etal2017} as 
\begin{equation}
\chi_{\nu}^2 = {\sum_{V,R,I}  \sum_{t_i \leq t_{pl}} { (m_{obs}(t_i) - m_{th}(t_i))^2 \over \sigma_i^2}} \times {1 \over (n_V +n_R +n_I -3)}
\end{equation} 
%\begin{wrapfigure}{l}{0.6\textwidth}
%\includegraphics[trim=-10 0 14 11,clip, width =0.6\textwidth, height= 6.5 cm]{den_compare_m.eps}
%\vspace{0.5em}
%\includegraphics[trim=-10 0 14 8,clip, width =0.6\textwidth, height= 6.5 cm]{den_compare_r.eps}
%\caption{Pre-SN density profile of a progenitor of ZAMS mass 13 $M_\odot$. The model without any enhanced mass loss and the one with shells of CSM outside have been shown in same plot for comparison. The radial profile of the central part of $\sim 4.5 M_\odot$ has been excluded from the plot on right side to show the CSM with better resolution.}
%\label{fig2}
%\end{wrapfigure}
% In SNEC, mass of the NS has been taken to be equal to Fe core mass and $^{56}$Ni has been evenly spread from outside the Fe core to the middle of He core. After boxcar smoothing this distribution gets modified with a tail extending upto the He core boundary.} 
Here $t_{pl}$ is plateau length which is $\sim$ 99 days \footnote{By fitting the V-band light curve with the function $y(t) = {{{-a_0} / {(1+{exp{{t-t_{pl}}\over w_0}})}} + (p_0\times t) + m_0}$ as proposed in \cite{Valenti_etal2016}, the best fit value of the plateau duration is $t_{pl}$= 98.77 days}, $m_{th}$ = $M_{th}$ + (distance and extinction correction),  $n_\lambda$ is no. of data points in a particular band. All sources of errors i.e. the photometric error in observational data, uncertainty in distance and extinction estimation have been included in $\sigma$. We have used the distance and interstellar extinction correction quoted in \cite{Rich2014}, \cite{Bose_etal2015}, \cite{Huang_etal2015} and \cite{Yuan_etal2016}. We find that for the set of values quoted in Richmond the best fitted model with CSM has ZAMS mass 12- 13$M_\odot$ and explosion energy $E_{exp}$ = $0.6\times 10^{51}$ergs, while the values of Yuan and Bose predict best fitted ZAMS mass = 13$M_\odot$ and a range of $E_{exp}$ = $0.6-0.8 \times 10^{51}$ergs. Huang has considered a larger value of extinction by taking the reddening of host galaxy M74 into account. Using their values we can pinpoint a ZAMS mass 13$M_\odot$ and $E_{exp}$ = $0.8\times 10^{51}$ergs. The explosion of models without CSM deviate from the data with a different distribution (Fig. \ref{fig3}), and with an error larger by a factor of $\sim$6-8 than explosions powered by the circumstellar medium (Table \ref{tab1}). The results have been cross-checked against the data from \cite{Valenti_etal2014} and \cite{Bose_etal2015} as well. It turns out that the model with CSM best fitted using Huang's values is visually closest to the data (Fig. \ref{fig4}). It strengthens the consideration of host extinction which is confirmed by recent analysis of massive star population around the site of explosion(\cite{Maund2017}). Although it was intended to fit upto the plateau only, since SNEC is not reliable beyond that region (\cite{MorozovaV_etal15}), the model with dense and nearby CSM fits the radioactive tail reasonably well in three bands. Also, the model predicts the peak magnitude in U and B bands correctly. On the other hand, the best-fit model without CSM hardly satisfies any feature of the light curve. This analysis is consistent with the spectroscopic observations of \cite{Bose_etal2015} where a weak ejecta-CSM interaction was inferred from the high velocity $H\alpha-H\beta$ profiles. This suggests that the progenitor of SN2013ej was surrounded by dense CSM and supports the suggestion of \cite{SmithN_etal2014} of heavy mass loss immediately prior to the SN explosion.   
\begin{table}[h]
  \begin{center}
  \caption{Overview of progenitor and explosion properties best-fitted for SN2013ej}
  \label{tab1}
  \begin{tabular}{c c c c c c c}\hline 
{\bf ZAMS} & {\bf Pre-SN} & {\bf CSM } & {\bf Pre-SN } & {\bf CSM } & {\bf Energy} & {\bf Goodness of Fit} \\
{\bf mass [$M_\odot$]} & {\bf mass [$M_\odot$]} & {\bf mass [$M_\odot$]$^1$} & {\bf radius [$R_\odot$]} & {\bf radius [$R_\odot$]$^2$} & {\bf [$10^{51} $ergs]$^3$} & {[\bf$\chi_{\nu}^2$]}  \\ \hline
 13 & 11.60 & 0.76 & 667 & 1650 & 0.6-0.8 & 0.0019-0.0028 \\ %\hline
 13 & 12.36 & ---  & 617 & --- & 1.0-1.2 & 0.0124-0.0227 \\ \hline
  \end{tabular}
 \end{center}
\vspace{5mm}
 \normalsize{
 {\it Notes:}\\
  $^1$ lost mass during enhanced mass loss rate in last ($\sim 2-3$)few years. \\
  $^2$ the external radius of the dense CSM formed in the last few years before explosion \\
  $^3$ asymptotic energy of the shock after breakout\\
  $^4$ $^{56}$Ni mass was kept fixed at 0.021$M_\odot$}
\end{table}

%\begin{wrapfigure}{l}{0.7\textwidth}
%\includegraphics[trim=10 12 0 35,clip, width =0.7\textwidth, height= 6.5 cm]{Error_map_CSM_Yuan.eps}
%\vspace{1em}
%\includegraphics[trim=10 14 0 35,clip, width =0.7\textwidth, height= 6.5 cm]{Error_map_Yuan.eps}
%\caption{Error maps of the models (left: with CSM, right:without CSM)of different ZAMS mass and explosion energy against V,R and I bands of SN2013ej (data: \cite{Yuan_etal2016}). The photometric error in observational data, errors in estimation of the distance of host galaxy and average galactic extinction (taken from \cite{Huang_etal2015}) have been included in the error analysis.} 
%\label{fig3}
%\end{wrapfigure} 
\section{Discussion}
\hspace*{1 cm} In the very late stage of post-main-sequence evolution of RSGs the mass loss rate can increase by several orders of magnitude (\cite{Quat2012}; \cite{SmithN_etal2014}; \cite{Moriya2014}). This can give rise to a dense and slowly moving circumstellar gas just outside the star. Once the shock comes out of the star it hits onto the CSM and deposits a larger fraction of its energy to the CSM which gets diffusively radiated and brightens the early optical light curves by degrading the adiabatic loss. \\ \\
\hspace*{1 cm} Here we have discussed how the same star can pass through two evolutionary tracks differing in a timescale much less than the Kelvin-Helmholtz timescale and end up as different pre-SN progenitors, although occupying roughly the same position in H-R diagram. The density, velocity and temperature profiles of hydrogen envelope of the stars with no huge mass loss history are substantially different from those of the progenitors surrounded by dense and compact CSM (Fig.\ref{fig2}); even though total mass (envelope + CSM) is same in both cases. So the rise and plateau of the light curve, which are functions of hydrogen profile are expected to be different. In these cases, although the pre-SN images cannot be distinguished, the light curves of explosion will differ. We note some particular differences in the presence of dense CSM: UV and optical bands are simultaneously bright in early phase, for a fixed shock energy the optical light curve is brighter at peak, flatter after peak and the plateau is longer. The dense and compact CSM behaves like an unbound extended part of the progenitor itself. The key difference between this scenario and usual II-n SNe is that the extended CSM in II-n is optically thin, while here the photosphere lies in the compact CSM for $\sim$40 days since explosion. In usual II-n the interactive phase lasts for $\sim$1 yr but in our case that is $\sim$ 2-3 days and the breakout happens when the shock has already traversed through the CSM. Radio and flash-ionization non-detection early on can constrain the radial extent of the CSM and confirm proximity and compactness (\cite{Yaron_etal2017}).  \\ \\  
%\hspace*{1 cm} We can extend this to II-L for different set of values of CSM parameters, and try to unify a subsection of II-P and II-L where early light curves are influenced by dense and compact CSM. CSM surrounded II-P/II-L SNe can be used to understand eruption and late-time outbursts in RSGs. For YSGs, BSGs and LBVs where the mass loss history is more varied, counting the effect of late stage evolution can lead to rich varieties of SNe as well.      


\begin{thebibliography}{}

\bibitem[Bose \etal\ 2015]{Bose_etal2015}
{Bose S. \etal\,} 2015,\textit{ApJ}, 806, 160, 18 pp.

\bibitem[Cardelli \etal\ 1989]{Cardelli} 
{Cardelli J. A., Geoferey G. C. \& Mathis J. S.,} 1989, \textit{ApJ}, 345:245-256

\bibitem[Cedr\'es \etal\ 2012]{Cedres} 
{Cedr\'es B., Cepa J., Bongiovanni \'A. \etal\,} 2012, \textit{A}\&\textit{A}, 545, A43

\bibitem[Chakraborti \etal\ 2016]{Chak_etal2016}
{Chakraborti S. \etal\,} 2016, \textit{ApJ}, 817, 22, 8 pp.

%\bibitem[Dessart \etal\ 2017]{Des_etal2017}
%{Dessart L., Hillier D. J., Audit E.,} 2017arXiv170401697D
	
\bibitem[Fraser \etal\ 2014]{Fraser_etal2014}
{Fraser M. \etal\,} 2014, \textit{MNRAS}, 439, L56-L60

\bibitem[Huang \etal\ 2015]{Huang_etal2015}
{Huang F. \etal\,} 2015, \textit{ApJ}, 807, 59 , 12 pp.

\bibitem[Maund 2017]{Maund2017}
{Maund J. R.,} 2017,	\textit{MNRAS}, 2017arXiv170401957M

\bibitem[Moriya \etal\ 2011]{Moriya_etal2011}
{Moriya T., Tominaga N., Blinnikov S. I., Baklanov P. V., Sorokina E. I.,} 2011, 
\textit{MNRAS}, 415, 199-213

\bibitem[Moriya 2014]{Moriya2014}
{Moriya T.,} 2014, \textit{A}\&\textit{A}, 564, A83

\bibitem[Moriya \etal\ 2017]{Moriya_etal2017}
{Moriya T. J., Yoon S. C., Gr$\ddot{a}$fener G., Blinnikov S. I.,} 2017arXiv170303084M

\bibitem[Morozova, Ott \& Piro 2015]{MorozovaV_etal15}
{Morozova V., Ott C. D., Piro A. L.,} 2015,
\textit{	Astrophysics Source Code Library,} record ascl:1505.033

\bibitem[Morozova \etal\ 2015]{MorozovaV_etal2015}
{Morozova V. \etal\,} 2015, \textit{ApJ}, 814, 63, 18 pp.

\bibitem[Morozova \etal\ 2017]{MorozovaV_etal2017}
{Morozova V., Piro, A. L., Valenti, S.,} 2017, \textit{ApJ}, 838, 28, 12 pp. 

\bibitem[Paxton \etal\ 2011]{Pax2011}
{Paxton B. \etal\ ,} 2011, \textit{ApJS}, 192, 3, 35 pp. 

\bibitem[Paxton \etal\ 2013]{Pax2013}
{Paxton B. \etal\ ,} 2013, \textit{ApJS}, 208, 4, 42 pp. 

\bibitem[Paxton \etal\ 2015]{Pax2015}
{Paxton B. \etal\ ,} 2015, \textit{ApJSS}, 220, 15, 44 pp. 

\bibitem[Quataert \& Shiode 2012]{Quat2012}
{Quataert E. \& Shiode J.,} 2012, \textit{MNRAS}, 423, L92 - L96

\bibitem[Richmond 2014]{Rich2014}
{Richmond M. W.,} 2014, \textit{JAAVSO}, 42, 333

\bibitem[Sanders \etal\ 2015]{Sanders_etal2015}
{Sanders N. E. \etal\,} 2015, \textit{ApJ}, 799, 208, 23 pp. 

\bibitem[Smith \etal\ 2011]{SmithN_etal2011}
{Smith N. \etal\,} 2011, \textit{MNRAS}, 412, 1512 

\bibitem[Smith \& Arnett 2014]{SmithN_etal2014}
{Smith N. \& Arnett D. W.,} 2014, \textit{ApJ}, 785, 82, 12 pp.

\bibitem[Valenti \etal\ 2014]{Valenti_etal2014}
{Valenti S. \etal\,} 2014, \textit{MNRAS}, 438, L101-L105

\bibitem[Valenti \etal\ 2015]{Valenti_etal2015}
{Valenti S. \etal\,} 2015, \textit{MNRAS}, 448, 2608-2616

\bibitem[Valenti \etal\ 2016]{Valenti_etal2016} 
{Valenti S. \etal\,} 2016, \textit{MNRAS}, 459, 3939-3962

\bibitem[Yaron \etal\ 2017]{Yaron_etal2017}
{Yaron O. \etal\,} 2017, \textit{Nature Physics}, 4025

\bibitem[Yuan \etal\ 2016]{Yuan_etal2016}
{Yuan F. \etal\,} 2016, \textit{MNRAS} 461, 2003 - 2018

\end{thebibliography}
\end{document}